# *Tracking galaxy evolution through low frequency radio continuum observations using SKA and Citizen-science Research using Multi-wavelength data*

**Ananda Hota** [1,2], **Chiranjib Konar** [3,2], **C. S. Stalin** [4], **Sravani Vaddi**[5], **Pradeepta K. Mohanty**[2], **Pratik Dabhade**[2], **Sai Arun Dharmik Bhoga**[2], **Megha Rajoria**[2], **Sagar Sethi**[2]

1. #eAstroLab , UM-DAE Centre for Excellence in Basic Sciences, Vidyanagari, Mumbai, India
2. RAD@home Astronomy Collaboratory, India (hotaananda@gmail.com)
3. Amity University Uttar Pradesh, Sector-125, Noida, Uttar Pradesh, India
4. Indian Institute of Astrophysics, Koramangla, Bangalore, India
5. National Centre for Radio Astrophysics, TIFR, Ganeshkhind, Post Bag no. 3, Pune, India

***Abstract:*** *We present a brief review of progress in the understanding of general spiral and elliptical galaxies, through merger, star formation and AGN activities. With reference to case studies performed with the GMRT, we highlight the unique aspects of studying galaxies in the radio wavelengths where powerful quasars and bright radio galaxies are the dominating subjects traditionally. Though AGN or quasar activity is extremely energetic, it is extremely short-lived. This justify focusing on transitional galaxies to find relic-evidences of the immediate past AGN-feedback which decide the future course of evolution of a galaxy. Relic radio lobes can be best detected in low frequency observations with the GMRT, LOFAR and in future SKA. The age of these relic radio plasma can be as old as a few hundred Myr. There is a huge gap between this and what is found in optical bands. The very first relic-evidences of a past quasar activity ("Hanny's Voorwerp") was discovered in 2007 by a Galaxy Zoo citizen-scientists, a school teacher, in the optical bands. This relic is around a few tens of thousand years old. More discoveries needed to match these time-scales with star formation time-scales in AGN host galaxies to better understand black hole galaxy co-evolution process via feedback-driven quenching of star formation. It is now well-accepted that discovery and characterisation of such faint fuzzy relic features can be more efficiently done by human than machine. Radio interferometry images are more complicated than optical and need the citizen-scientists to be trained. RAD@home, the only Indian citizen-science research project in astronomy, analysing TIFR GMRT Sky Survey (TGSS) 150 MHz data and observing from the Giant Meterwave Radio Telescope (GMRT), was launched in April 2013. Unique, zero-infrastructure zero-funded design of RAD@home as a collaboratory of 69 trained e-astronomers is briefly described. Some of the new-found objects like episodic radio galaxies, radio-jet and companion galaxy interaction, radio galaxy bent by motion of the intra-filament medium in a Mpc-scale galaxy filament etc. are briefly presented as demonstration of its potential. Citizen-science has not only opened up a new way for astronomy research but also possibly the only promising way to extract maximum science out of the Big Data in the SKA-era. This possibly can convert the Big Data problem in to a prospect. Citizen-science can contribute to the knowledge creation in never-seen-before speed and in approach. As it is based on Internet, it can provide an equal opportunity of academic-growth to people even in the under-developed regions where we always need to put our optical and radio telescopes. This can liberate the research-activity of city-based research-institutes out of the four brick walls and alleviate various socio-economic and geo-political constraints on growth of citizens educated in undergraduate-level science but located in remote areas.*

*1. Introduction:*
Traditionally, powerful radio galaxies and quasars have dominated the observations of radio continuum emission from the sky. Until the sensitive XMM-Newton and Chandra X-ray observations were made, radio lobes, located far outside stellar emission, had almost no connection with observations other than the radio itself. Ghost-cavities, seen in X-ray emission lacking any radio counterpart in the Very Large Array (VLA) data but later found to be filled in with low-frequency radio emission from the Giant Meterwave Radio Telescope (GMRT), made the very first and direct radio-X-ray connections. This suggested interaction between thermal and non-thermal plasma on scales larger than galaxies (McNamara et al. 2001, Randall et al. 2011). Specially taking due advantage of 610-235 MHz dual frequency observations of the GMRT, Giacintucci et al. (2011) found these cavities to be filled with diffuse steep spectra radio emission which can be naturally accepted as a relic-evidence of a past AGN (radio galaxy) activity. Double-Double radio galaxies (DDRG) are most-convincing evidences of episodic activity of AGNs, in particular the radio loud cases (Reynolds & Begelman (1997), Saikia & Jamrozy (2009)). Similar to triple double radio galaxies, suggesting three distinct AGN-radio-jet episodes in B0925+420 (Brocksopp et al 2007), Speca (Hota et al. 2011), and J1216+0709 (Singh et al. 2016), three pairs of X-ray cavities have also been seen in the galaxy NGC5813 (Randall et al. 2011).

Remarkable discovery in optical bands of what is now known as "Hanny's Voorwerp" is the optical counterpart of such relic-evidences of past AGN/quasar activities (Lintott et al. 2009). Immediately after this discovery in 2007, citizen-scientists have reported a dozen of such Hanny's Voorwerps (Keel et al. 2012a). Optical spectroscopy of these faint fuzzy coloured ionised gas clouds have revealed that they have been ionised by radiation from quasar in the nearby associated galaxy which has suddenly become dormant/dead since last ten thousand to one Myr (Schawinski et al. 2010, Keel et al. 2012b). While these optical emission line relics are nearly a few tens of thousand years old, fossil or relic-plasma from radio galaxy lobes, with no jets and hotspots, can be a few hundred Myr old (e.g. Murgia et al. 2011, Hunik & Jamrozy 2016, Konar et al. 2013, Giacintucci et al. 2012, Hota et al. 2011). Thus, the radio band has the capability of tracking an order of magnitude longer history of AGN activity than optical. However, plots like colour-magnitude diagrams and population synthesis modeling using UV and optical photometric data, can typically track a last major episode of star formation or residual star formation in transitional or post-starburst galaxies from a few hundred Myr to upto billion years (Bruzual & Charlot 2003, Schawinski et al. 2007; Kaviraj et al. 2011). One billion year is also a typical time-scale for a merger between two large spiral galaxies. Although quasar-wind, galactic wind or superwind and radio jet feedbacks are powerful and their impacts are long-lasting, the feedback outflows are very short events in the life of a galaxy (Heckman et al. 1990, Veilleux, Cecil & Bland-Hawthorn 2005 , Springel, Di Matteo & Hernquist 2005, Croton et al. 2006, Hopkins et al. 2006, Fabian 2012, Kormendy & Ho 2013). Hence, to find a direct evidence and investigate the role of AGN jet/wind feedback during and after galaxy merger, it is necessary to discover new targets and push back the time-scale of AGN-relics, from a few 100 Myr up to a billion yr, if possible. Sensitive, low-frequency and wide field imaging studies with the Square Kilometre Array (SKA) will play a vital role in such a discovery. Some leading examples have been found using the GMRT  (Speca; Hota et al 2011, J021659-044920; Tamhane et al. 2015) and many will be found soon with the TIFR GMRT Sky Survey (TGSS) at 150 MHz and future Low Frequency Array (LOFAR) surveys. Such studies together with sensitive low frequency observations with the SKA, will be an important step in revealing the complete evolutionary history of galaxies in terms of galaxy merger, star formation, AGN activity, black hole merger, and AGN- jet/wind feedback processes.  Since AGN activity peaks at redshift of 2-3 (Nesvadba et al. 2008), while current low frequency detections are limited to the nearby Universe, there is a need to push the sensitivity to reach the quasar-era in detecting relics or remnants lobes around these AGNs.

Due to the high sensitivity and large field of view, the data production rate of the SKA and other future optical-IR survey telescopes will be extremely high. This is popularly known as BIG DATA problem, and it is not just a technological problem but also a human-problem for how to make sense of the data and convert information to knowledge (Michael Nielsen 2011, Marshall, Lintott & Flether 2015). One potential way out is citizen-science research which is also referred to as crowd sourcing, networked science or public-participation in research. It seems critical to employ citizen-scientists, up to million members, to locate faint fuzzy fossil evidences of AGN activity, in any wave band and in any database available on the web. A few leading examples discovered by citizen-scientists will help the professional astronomers for making targeted deep and multi-wavelength follow up investigations for better understanding of black hole-galaxy co-evolution through merger, feedback and alternate paths (see a comprehensive review by Kormendy & Ho (2013).

In this article we consider galaxies in general, which include radio quiet or normal galaxies and several kinds of transitional galaxies which may or may not contain AGNs, and may or may not have radio jets and lobes of luminosity to be called radio-loud. There is a natural need to study them, being unbiased towards the presence of starburst or AGN in it, as it is well-known that AGN-activity is a transient phase in a galaxy. A galaxy either was an AGN once upon a time, or will be an AGN in future. Similarly, nearly equal fraction of star forming galaxies are seen in radio-faint and radio-bright early-type galaxies (Vaddi et al. 2016). This approach is justified from another perspective as well. Radio galaxies are mostly associated with only elliptical galaxies. This has led to the widely accepted explanation that spiral galaxies merge leading to formation of eliipticals and in the process their central black holes also merge to become super massive and fast spinning which is required to launch large relativistic radio jets (Blandford & Zjanek 1977, Heckman et al. 1986, Wilson & Colbert 1995, Sikora, Stawarz & Lasota 2007, Hopkins et al. 2008). With the recent discovery of several spiral-host large radio galaxies a possible new window in understanding galaxy evolution is opening (Hota et al. 2011, 2014, Bagchi et al. 2014, Mao et al. 2015, Singh et al 2015). It is very likely that these spiral-host large radio galaxies have avoided merger with other large galaxies and with clusters and have been growing in isolation, through accretion from filaments in the cosmic web. These are like 'living fossils' or messengers from the early Universe which are possibly the first massive spirals with supermassive black holes in it. In the last part of the article, Sec. 4 and 5, we have reviewed the citizen-science research in astronomy with special reference to the modified citizen-science research, RAD@home, being done in India using GMRT observations and TGSS data. We briefly describe our future plans, the necessity and prospect of doing citizen-science as large web-based professional-amateur collaboration during the SKA era.

*2. Transitional galaxies as Laboratory for Evolution study:*
We shall focus on some special type of galaxies, ignoring standard spiral and ellipticals. We consider cases where the phase of the galaxy is transient, such as galactic wind (superwind), extreme starburst, merger, AGN-jet or quasar-wind and/or odd natured blue-elliptical, red-spiral, spiral-host radio galaxy. These are selected because in this transitional phase the physical processes that bring dramatic changes and drive fast galaxy evolution can be caught while in the act of transformation. With the detailed investigation of such targets using new telescope capacity, the knowledge of the field can be easily and significantly expanded. It will be described later in detail that although the targets of interest are these transitional galaxies, the multi-wavelength citizen-science approach that we follow (RAD@home) will analyse every galaxy in the wide field of view of SKA.

## 2.1 Spiral galaxies with radio bubbles:

Radio observations of spiral galaxies show synchrotron and free-free emission from cosmic ray particles originating from supernova remnants and ionised gas present in the star forming regions, respectively (Condon 1992). At higher frequencies, sub-mm and far-infrared (FIR), the thermal (black body) emission from star forming region dominates. The tight correlation between FIR and radio flux densities of normal star forming galaxies is one of the long-standing puzzles in extra-galactic astronomy (Condon 1992). While cosmic ray particles accelerated in the supernova remnants diffuse out and create wide spread long-lasting synchrotron emission in both the spiral arms and inter-arm regions, free-free emission from star forming regions is more localised in the spiral arms and short-lived. Low frequency radio imaging with the GMRT at 333 MHz was used to understand radio-FIR correlation by Basu et al (2012). They found the inter-arm region to have different radio-FIR slope than the spiral arm region which is attributed to propagation of low energy (~1.5 GeV) and long-lived (nearly 100 Myr) cosmic-ray electrons at low radio frequencies. Nearby galaxies like M82 and NGC253 clearly show radio emission much beyond the stellar distribution or the star forming disk and it is attributed to the diffusion of synchrotron emitting cosmic ray particles into the halo of the galaxy (Seaquist & Odegard 1991, Carilli et al. 1992). However, in some cases well-defined bipolar jets and bubbles of radio emissions are seen, with no direct correspondences to the star forming regions and extending out of the stellar disk. Two well-known examples are Circinus galaxy (Elmouttie et al. 1998) and NGC3079. GMRT low frequency observations of NGC3079 have shown not only the correspondence with the X-ray and Hα outflow but also signs of episodic nature of the central AGN, with radio jet. This jet has probably supplied the relativistic plasma to the radio bubble (Irwin & Saikia 2003). Radio observations of large samples of spiral galaxies with such radio structures extending away from the stellar disks, has been a puzzle as to which of the following two processes is dominant; starburst-driven superwind or AGN-jet, though small in size and short-lived (Baum et al. 1993, Colbert et al. 1996, Galimore et etl 2006, Hota & Saikia 2006). A sample of ten radio bubble galaxies compiled by Hota & Saikia (2006) shows that every object has an AGN at the center, suggesting the AGN to be responsible for supplying the synchrotron plasma and may have inflated the bubble with or without the dynamical influence from the starburst-driven superwind outflow. GMRT data on such a spiral galaxy sample, Seyferts with radio jet/bubble outflows, have already been collected. As these sensitive observations are in an unexplored low frequency range, it has the potential of revealing many of these Seyferts to have episodic radio lobes. Mrk 6 has been the only Seyfert with episodic radio jets or lobes but GMRT 610 and 325 MHz observation have found possible relic lobe emission beside NGC4235 making it possibly the second Seyfert with episodic radio jets (Kharb et al. 2016).

**2.1.1 NGC6764:** We briefly discuss here the case of NGC6764 which has been investigated in detail with the GMRT, VLA and Chandra X-ray telescope. Hota & Saikia (2006) showed that the radio bubble of this barred spiral galaxy has an associated ionised gas outflow, seen in Hα and [NII] spectral-line imaging and long-slit kinematics study. The radio structure close to nucleus is seen orthogonal to the large-scale orientation of the radio bubble, likely suggesting that the nuclear jet-like outflow is disrupted by dense molecular gas in the central region. GMRT observations also show HI gas to be outflowing at nearly 120 km/s, as seen in blue-shifted absorption line. Further support to the outflow process is also seen in cold molecular gas observations (Leon et al. 2007). Follow up study with the Chandra X-ray observatory demonstrated that the X-ray emitting bubbles are co-spatial with the radio-bubbles (Croston et al. 2008). This study rules out the starburst wind to be generating the X-ray and favours the radio-bubble to be hitting the ISM and shock-heating to create the X-ray bubble. Surprisingly, the total energy stored in the AGN-heated hot gas bubbles is nearly $10^{56}$ ergs and that is comparable to energetic impact of low-power radio galaxies such as Centaurus A hosted in an elliptical galaxy. This demonstrates the possibility of dramatic impact of short-lived, frequent, fast-precessing, radio jet/bubble outflow from typical spiral galaxies and its wide implications on quenching of star formation. It was also noticed that the dynamical (expansion as well as jet precession) time scale of 12-21 Myr of the bubbles (jet launched 21-12 Myr ago) roughly coincide with a break in the two episodes of nuclear star formation history (50-15 Myr and 5-3 Myr ago) of the galaxy (Kharb et al. 2010). If this match of time-scale can be supported with further observations for a causal connection, it is one of the beautiful cases of quenching of star formation by the feedback from low-power radio jets in spiral galaxies.

*2.2 Merger Remnant Luminous Infrared Galaxies (LIRGs):*

Merger of two gas-rich spiral galaxies lead to extremely high density of molecular gas and dust in the central region of the merger-remnant (Mirabel & Sanders 1996). This triggers high rate of star formation (Barnes & Hernquist 1991). The UV photons from hot, young, and massive stars get absorbed by the surrounding dust and make the remnant Hyper/Ultra Luminous Infrared Galaxies (LIRGs). Gas and stars in the tidal tails, continue to rain-down from the orbital plane of the merger on to the remnant for about a billion year (Springel et al. 2005). These tails remain bright in UV and blue optical light, due to continuation of the star formation, even after the merger of the nuclei. Infrared and radio remains the only bands that can peep inside the dust-vails and reveal what has been happening at the centre. At the centre, while two galactic nuclei are about to merge, each nucleus may go through compact star burst and/or quasar activity (with or without radio jet). Two well-known examples are NGC6240 and Arp220 (Mirabel & Sanders 1996 and references there in). These central activities lead to powerful superwind outflows eventually removing significant fraction of gas and dust from the central region to give birth to a classical UV- and optically-bright quasar. Herschel-PACS observations of Hydroxyl (OH) molecules in far-IR spectra of ULIRGs, showed massive outflow of molecular gas with velocities, in some cases, exceeding 1000 km/s and mass outflow rates up to 1200 solar mass per year (Sturm et al. 2011). These outflow rates are several times larger than the star formation rate and, suggest clear and severe impact of negative feedback on the future of ULIRGs. Similarly, at sub-mm wavelength observations using IRAM Plateau de Bure
 Interferometer (PdBI)
have established massive cold gas CO(1–0)
 outflow at rates comparable or higher than the star formation rate (Cicone et al. 2014). Recent ALMA observations provide substantial evidences for cold molecular gas outflow to be affecting future of LIRGs (e.g. Sakamoto et al. 2014, Emonts et al. 2015, Aalto et al. 2016, ). Molecular gas outflow, driven by AGN radio-bubble, has also been observed in an early-type galaxy with no detectable sign of past galaxy interaction (Alatalo et al. 2011). Thus such outflows, in multiple episodes, can ultimately lead to formation of gas-poor elliptical galaxies to launch, without hindrance, large radio jets and lobes, known as classical radio galaxies. Many LIRGs with close pair of nuclei and having massive outflow of HI, molecular gas (CO) and dust may represent beautiful example of LIRGs as transition phase from gas-rich starburst merger to standard quasars or elliptical-host radio galaxies.

Taking due advantage of dual-frequency 610-235 MHz observations, several samples of LIRGs have been observed with the GMRT (Clemens et al. 2010). This dual-frequency is critical in determining the spectral shape, as observations done in different epoch have a possibility of being affected by the variability of the compact radio source if the central source is an AGN ( e.g. Mrk 231; Condon et al. 1991). Radio continuum can look through the dust but at lower frequencies it gets absorbed by the thick ionised gas (free-free absorption) at the central region of LIRGs, giving us a unique opportunity to measure ionised gas amount, otherwise not possible by optical observations (Condon et al. 1991). Clemens et al. finds that the lifetime of ionised gas is an order of magnitude more than the HII-regions, and in sources showing free-free absorption the star formation rate is still increasing with time. For multi-frequency SKA observation. A sample of LIRGs, arranged in the order of distance between two nuclei, ignoring projection, can serve as a beautiful project in investigating evolution of hidden nuclear activities of LIRGs. This can investigate evolution of ionised gas, star formation rate, supernova rate, interplay and co-evolution of starformation and AGN activity. Note that the fraction of AGNs in a LIRG sample increases systematically with luminosity and luminosity increases with the decrease of inter-nuclei distance. One can speculate that sensitive SKA observations may reveal episodes of feedback not only in low-frequency imaging but also in HI-gas, both in emission and absorption. Here we present result from such a detail study of a merger remnant NGC1482 done with the GMRT.

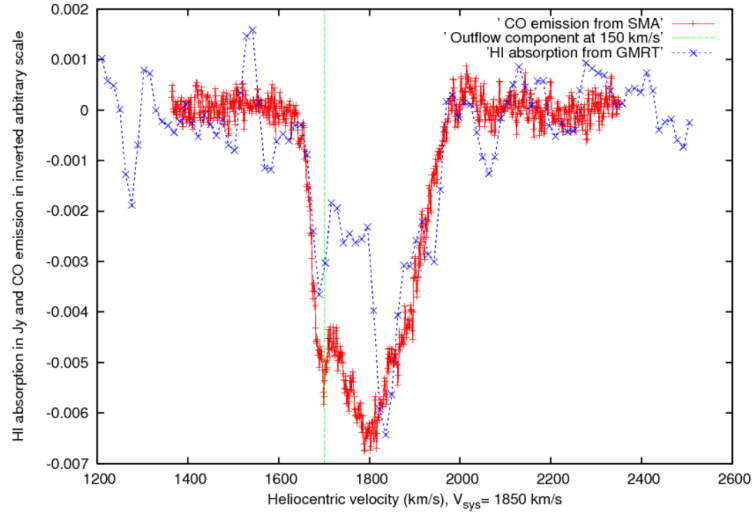

*Fig. 1: The HI-absorption line (blue) is superposed on the CO (J=2-1) emission (red) from the central region of NGC1482. The $^{12}CO$ J=2-1 emission, from the SMA, has been scaled and inverted for compactness and ease of comparison with the HI-line from the GMRT. Notice the CO emission component coincide with the blue-shifted HI-absorption at 150 km/s (Hota & Saikia 2005, Hota, Espada, Matsushita et al. 2016)*

*2.2.1 NGC1482:* *This* is probably the only post-merger early-type galaxy with a Superwind outflow in the nearby Universe within 30 Mpc, making it the ideal target to understand the process of evolution through merger and feedback (Veilleux & Rupke 2002, Strickland 2007). This unique target has been studied with both GMRT and VLA in complementary frequency ranges both in radio continuum and HI (Hota & Saikia 2005). This nearby early-type (S0/a) galaxy shows clear signs of past merger, in the form of tidal tails, in deep optical images. Supplementing this, HI observations also show two large (~60 kpc) tidal tails (Omar & Dwarakanath 2005) and near-IR observations show two nuclei (separation ~360 pc) at the central dusty region. Radio continuum observation is the only band to show the star forming disk at the base, encompassing both the nuclei. Launched from the base is a perfect hourglass-shaped bi-conical superwind outflow. The outflow is seen both in [NII]/Hα line ratio map and X-ray images from Chandra. Radio observations, on the basis of compactness, spectral index and variability, suggest the western nucleus to be an AGN and possibly has a tiny radio jet, seen in high-frequency VLA images (Hota & Saikia 2005). All phases of gas, hot (Chandra observation), warm (Hα), neutral (HI) and cold molecular gas ($^{12}CO$ J=2-1), have been observed to show evidences of mass outflow. Intriguingly, from the nuclear region, blue-shifted HI-absorption line, observed with the GMRT, and blue-shifted CO-emission line, observed with the Sub-Milimetre Array (SMA), show similar outflow velocity of ~150 km/s (Fig. 1, Hota & Saikia 2005, Hota, Espada, Matsushita et al. 2016 in preparation). This is observed at a stage while the nuclei are yet-to-merge ( separated by ~360 pc and having observed radial velocity difference of ~200 km/s in CO line emission) and it shows mass loss and decline of star formation. The UV images from GALEX suggest that the galaxy had a major episode of star formation nearly 300 Myr ago and only small region around the dust lane contains stars younger than 150 Myr. Whereas, the dynamical time-scale of the Superwind is very young, launched

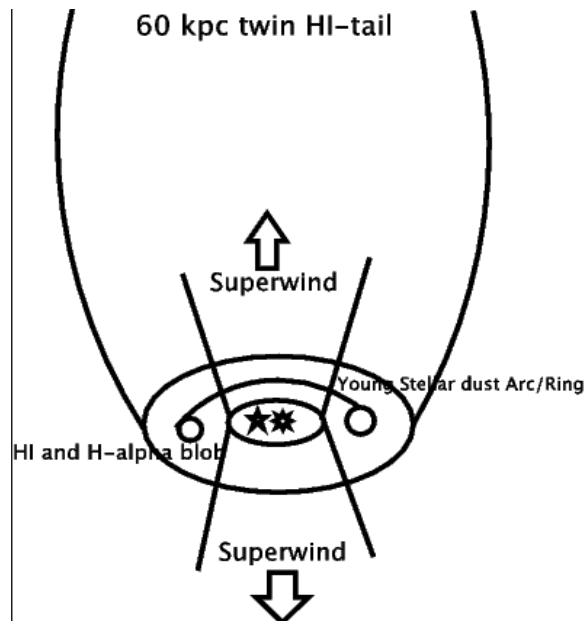

*Fig. 2: Big-picture summary for NGC1482 the merger-remnant, dual-nuclei, infrared luminous, early-type galaxy with superwind outflow (Hota & Saikia 2005)*

just ~6 Myr ago (Veilleux & Rupke 2002). Thus, the current episode of outflow cannot be the cause of the decline of star formation. Hence, multiple episodes of such mass losses, specially molecular gas, are needed to eventually lead to quenching of star formation and, possibly a red-and-dead early-type galaxy. Summary of NGC1482 observations has been presented in a schematic in Fig 2 with not-to-scale display of tidal tails, nuclear star formation, binary nuclei and superwind. NGC1482 presents a beautiful example for models of galaxy evolution through merger and galactic-wind feedback (Veilleux et al. 2005). This also serve as an example where radio observations can not only reveal the dust-hidden nuclear starburst and/or AGN as the power source but also lead kinematic evidences of outflow showing feedback processes in action for transforming the remnant galaxy.

**2.3 Post-merger and post-starburst radio galaxies:**
There is a huge observational gap between the gas-rich star forming merger-remnants with buried radio-bright AGNs and gas-poor ellipticals with sub-galactic young or giant fully evolved radio jets and lobes. No observational study exist to track down the time sequence of a gas-rich merger-remnant to evolve to a gas-poor elliptical. It is likely that the radio-jets indeed lose their significant amount of energy in working against the dense ISM to push out and travel far, nearly a few 100 kpc to be finally identified as a standard radio galaxy. Why would one expect such a sequence? As mentioned before, the distance between the nuclei (ignoring the projection) can serve as a proxy for time since the beginning of the merger. Indeed, review on LIRGs by Mirable & Sanders (1996) presents that the IR-luminosity of merger-remnant LIRGs increases with decrease of separation between the two merging nuclei and furthermore the relative fraction of AGNs (over compact starburst) burried inside remnants also increases with increasing IR-luminosity. This suggest that with the passage of time, the star formation declines and AGN activity peaks. High resolution radio imaging studies with Very Long Baseline Interferometry (VLBI), have found burried AGNs with tiny radio jets and lobes in several LIRGs (Lonsdale et al. 2003). Could the rise of AGN activity be the cause of fall of the starburst, through episodic feedback processes? With the arrival of Atacama Large Millimeter/submillimeter Array (ALMA) several remnants are now known to show massive molecular gas outflow, which can make the remnant deprived of fuel for future star formation. This starburst-superwind or AGN wind/jet driven feedback could be the reason, but relic/fossil-evidence of any kind in any reasonable sample size is still missing. This could have been nicely seen, if the old relativistic plasma lasts longer than we currently observe (a couple of hundred Myr only). Hence, both compact young and old (relic lobe) radio galaxies in merger remnant post-starburst galaxies are important targets for detailed investiations.

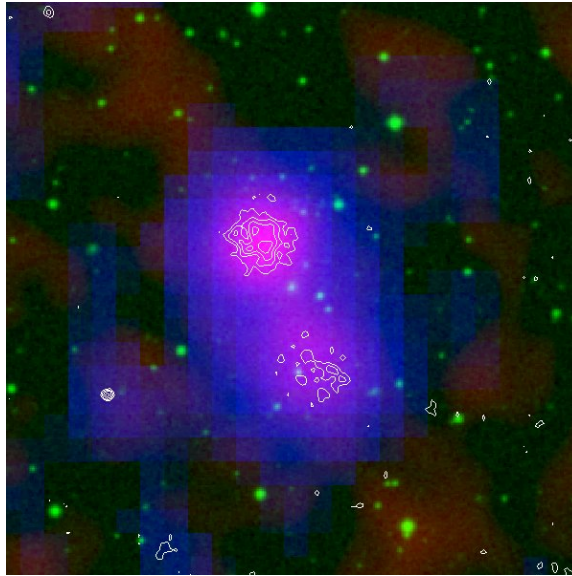

*Fig. 3:Radio-optical image of SDSSJ 161910.4+064223. Such Red (GMRT, TGSS ADR1, 150 MHz, b ~20''), Green (DSS2red), Blue (VLA, NVSS, 1.4 GHz, b=45'') and Contour ( VLA, FIRST, 1.4 GHz, b=5'') or RGB-C) multi-wavelength images are prepared using NASA-Skyview and analysed by RAD@home citizen-scientists or e-astronomers. The lack of hotsopt and back flow suggest the post-starburst galaxy to have a dead/relic radio lobe as well.*

**2.3.1 Post-starburst galaxies with relic radio lobes:** To support the models of feedback-driven quenching of star formation, there was a search of relic-radio lobes in post-starburst E+A galaxies by Shin, Strauss & Tojeiro (2011). These are the galaxies where star formation has abruptly been stopped with in last one billion year and jets from AGN could be the cause. Simply with the use of archival NRAO VLA Sky Survey (NVSS), Faint Images of the Radio Sky at Twenty-cm (FIRST) and *Sloan Digital Sky Survey (SDSS)* data, they have found only 12 out of total 513 post starburst galaxies to be bright in radio above 1 mJy. Among the dozen radio-detected sources only two are large radio sources where the lobes are diffuse and without any hotspot (Fig. 3). The radio spectrum also shows possible break, which will help estimating the age of the radio plasma. As the search was done in 1400 MHz the chance of detecting relic radio emission, possible relic evidence for quenching of star formation, is naturally low. We have already initiated a project to follow these objects with low-frequency observations with the GMRT. The same search with the newly available TIFR GMRT Sky Survey (TGSS) Alternative Data Release-1 (ADR-1; Intema et al. 2016) at 150 MHz is also underway. Radio spectral aging analyses can be combined with stellar population synthesis models to examine evolution of these two activities. In a reasonably large sample, with wide-field coverage and targeted deep low-frequency observations of post-starburst galaxies, it can be investigated, if indeed the AGN-activity could have preceded the decline of star formation. It may not be possible to detect relic lobe evidences for many galaxies in the quasar-era but should be possible at least to find examples in the nearby Universe demonstrating the process in action. A nice example of AGN-feedback in a post-merger starforming galaxy (UV-bright) with a very young radio jet feedback, NGC3801 popularly known as Cosmic Leafblower galaxy, is described below.

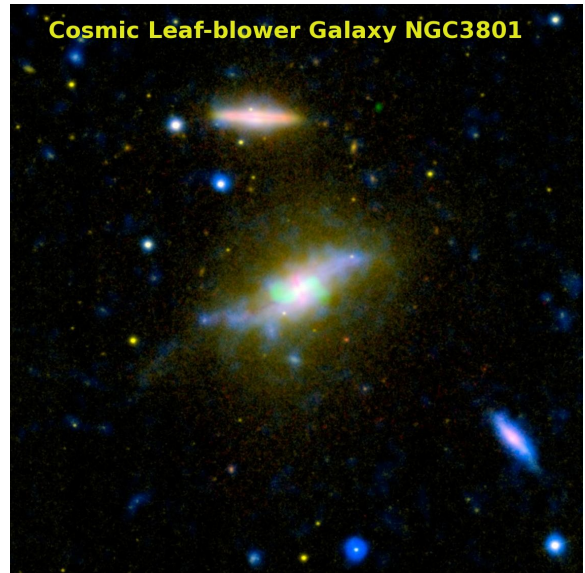

*Fig. 4: Multi-wavelength image of Cosmic Leaf-blower galaxy NGC3801 based on Hota et al. (2012). It shows SDSS optical image in yellow, Spitzer 8 micron in red, GALEX near UV in blue and VLA 1400 MHz radio emission in green.*

**2.3.2 NGC3801 Feedback caught-in-the-act:** Based on data from Hota et al. (2012) multi-wavelength (UV, optical, far-IR, radio) image of NGC3801 is presented in the Fig. 4 and a schematic to describe the possible evolutionary sequence is also presented in Fig 5. SDSS data shows NGC 3801 to be an optically red ($u' - r' = 3.08$) early-type galaxy with clear signs of past merger such as tidal tail, boxy and S-shaped optical isophotes and complex structure in dust. However, the UV images from GALEX telescope shows young star forming regions arranged in an S-shaped structure covering the whole galaxy (seen blue in Fig. 4). Stellar population synthesis modeling, using UV data, suggested that the major episode of star formation has already been over about 500-100 Myr ago. Thus the galaxy has similarity with *blue-ellipticals or post-merger post-starburst* galaxies. Interestingly, the galaxy has an young AGN, a S-shaped mini-radio lobes (~9 kpc), buried inside the stellar distribution. Spectral aging calculation in radio, suggest the radio lobes to be just 2-3 Myr old. Thus, NGC3801 has been caught at a crucial junction where following the merger of two spiral galaxies (a billion year time-scale) the star formation has declined (since ~500 Myr) and a young AGN radio-lobes have just erupted (2-3 Myr ago). Chandra X-ray observations by Croston et al. (2007) show two shock shells surrounding the radio lobes expanding at a speed of ~850 km/s into the ISM of this galaxy. HI and molecular gas (CO) observations and optical spectroscopy from HST, clearly demonstrate the galaxy to be a Kinematically Decoupled/Distinct Core (KDC) where the inner 5 kpc rotate around the major axis and the outer 10-20 kpc disk, both gas and stars, rotate around the minor axis. This KDC structure may be contributing to the bending of the mini radio lobes in to S- or Z-shaped structure. Hubble Space Telescope (HST) data and VLA data show outflow of ionised gas (up to 630 km/s) and neutral atomic hydrogen gas (HI) from the central and radio eastern lobe region, respectively (Hota et al. 2009). Interpreting data from varieties of wavelengths we suggested that the radio jets and its associated shock shells will hit the outer dust and gas-rich star forming region in the next 10 Myr (Hota et al. 2012). This time-scale is like a blink compared to a single rotation period of the gas in the galaxy which is nearly 330 Myr. A simplified sequences of events are speculated in the Fig. 5 for this unique AGN-feedback caught-in-the-act.

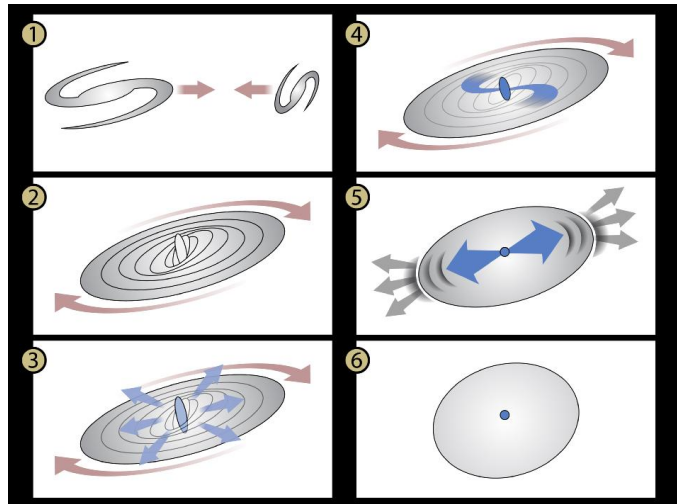

*Fig. 5: Schematic of the evolutionary sequence of NGC3801 which is based on Hota et al. (2012). It shows a speculated sequence of events as shown in the numbered panels. 1. Merger between two unequal mass spirals around a billion year ago 2. formation of the KDC in the remnant 3. possible starburst/quasar-wind-driven outflow around 500 Myr ago since when the star formation declined 4. Birth of a twin-radio jet, nearly 3 Myr ago, expanding orthogonal to the central gas disk and being bent by the outer rotating gas disk. 5. Representation of the shock waves and outflows driven by the radio lobe which can reach the outer gas disk in next 10 Myr. 6. transformation of the remnant in to a red-and-dead early-type galaxy.*

This galaxy has several similarity with Cen A, except that old diffuse radio emission outside the stellar light distribution is not seen, if any. Study can be carried out on a sample of post-merger post-starburst galaxies with such sub-galactic scale young radio sources to understand impact of the jet feedback on the galaxy itself and hunt of diffuse relic plasma, possibly from previous jet episodes, using low-frequency observations. GMRT observations on NGC3801 has already been carried out to look for HI signatures of previous episode of feedback, which may have happened earlier to this ejection of radio lobes which are only 3 Myr old. As the galaxy looks optically red and star formation declined nearly ~500 Myr ago, there is a need to look for much older episode of feedbacks in any possible observation mode. Next, we present an even more intriguing case of a early-type-early-type dry merger with two distinct episodes of radio jet feedbacks which interact with the ISM of the companion galaxy. This unique source was discovered by RAD@home citizen scientist and has been named RAD-18 (Hota et al. 2014)

**2.3.3 Dry merger episodic feedback radio galaxy RAD-18**:

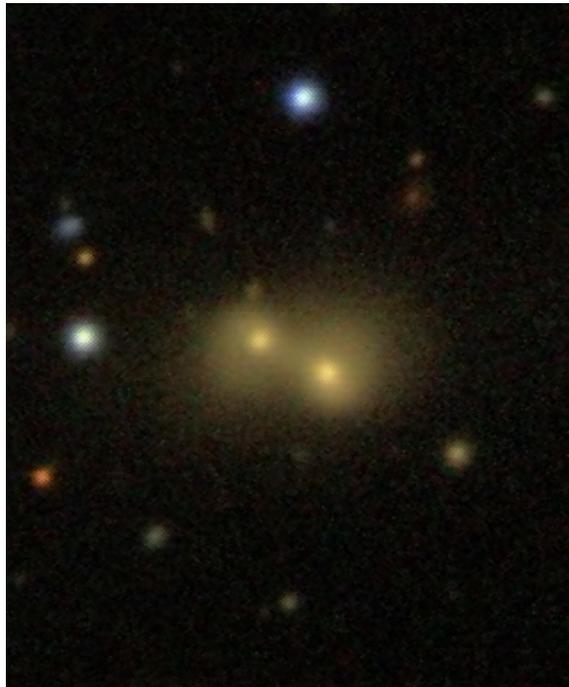

*Fig. 6: SDSS colour image of early-type early-type dry merger RAD-18. Note the fan-shaped stellar tidal tails which seem rotating clock-wise and roughly in the plane of the sky (Hota et al. 2014).*

Fig. 6 presents SDSS image of RAD-18 (RA: 13 30 10.3 Dec: -02 06 18 z=0.086), a case of dry merger which also shows distorted radio lobes (Fig. 7). This was identified from the DSS and VLA FIRST survey (Hota et al. 2014). Overlay of SDSS image and FIRST image (green contours) shows the western galaxy to be the host of the classic double radio lobe larger than the optical host galaxy. In addition to this double lobe structure, an extended feature is seen in the north and much fainter diffuse emission in the south west region. With the release of TGSS ADR-1 by Intema et al. (2016) the 150 MHz (b = 25") emission of the south-western (SW) faint diffuse emission became very clear which was present in the poor resolution NVSS (b = 45") image but was missing in high resolution FIRST (b = 5") image, both on 1400 MHz. This is likely a relic radio plasma from an earlier episode of AGN activity. However, the NE counterpart of this relic emission at a similar distance from the host galaxy is missing in the TGSS image. If we hypothesize that in an earlier episode of AGN activity, expansion of the SW lobe was undisturbed while the NE lobe was disturbed by the eastern early-type galaxy, many features can be explained. The interaction of the eastern galaxy and the previous episode of the jet-lobe (moving NE) could cause the radio plasma to be seen closer to the host and deflected towards the north. The optical galaxies change their position in the orbital plane of the merger. Evidence for the clockwise motion can be seen in the fan-shaped faint stellar tidal-tails seen in both these early-type galaxies (Fig. 6). The western galaxy moves southward and the eastern galaxy (and the tidal tail) moves northward. This northward motion, can be an additional cause for deflection of the radio plasma of the previous episode of jet on the NE side. Detailed radio structures can also be explained as the revolving galaxies and their associated ISM and tidal tails could compress and sweep the diffuse relic plasma of the earlier episode. Specially, the eastern galaxy pushing the NE relic lobe is likely creating the northern extended feature seen in the FIRST image. This shift in the location of the host galaxies, could have led to the relatively undisturbed expansion of the new radio lobes for the recent episode leading to a classic double lobe structure (FIRST image green contours). Higher resolution low frequency images and spectral index images at the resolution of FIRST would be useful in testing the above hypothesis.

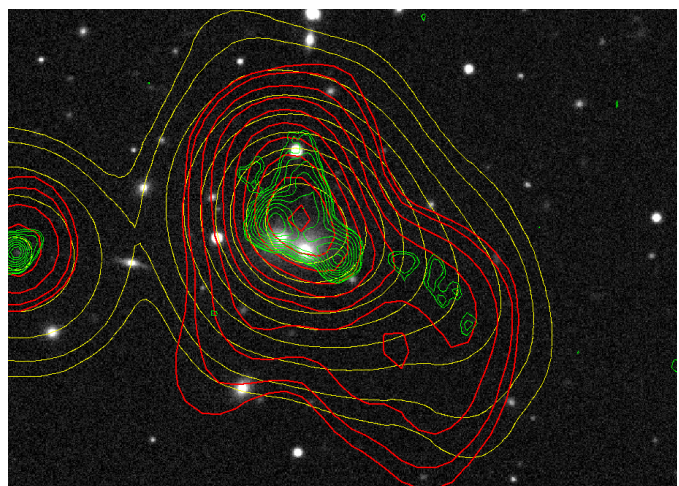

*Fig. 7: Multi-wavelength image of RAD-18. SDSS optical image in grey-scale is superposed with TGSS in red, NVSS in yellow and FIRST in green contours. Compact radio source on the left (east) is unrelated to RAD-18 (Hota et al. 2014).*

RAD-12 (RA 00 43 00.6, Dec -09 13 46, z=0.076) is yet another intriguing case discovered by the RAD@home (Hota et al. 2014). Here the radio jet from the eastern galaxy, directed towards the western companion at same redshift, is seen brightened up and has experienced lateral expansion. We speculate that the radio jet is interacting with the inter-stellar medium of the companion causing the jet to brighten up and stop its forward motion. This is unlike the bridge radio emission seen between two interacting spiral star forming galaxies (e.g. Taffy Galaxies; Condon et al. 1993). Here, both the galaxies are yellow in SDSS colour and structurally too look early-type. Deep observations to detect the eastern side of the jet is in plan.

It is amazing to notice that, in a fraction of the galaxy interaction time-scale and in spite of being gas-poor dry galaxy merger, the brighter western nucleus of RAD-18 could create two distinct episodes of radio jet and lobes. As seen in 3C321, popularly known as the Death Star galaxy (Evans et al. 2008) and Minkowski's object (Croft et al. 2006), the radio jet hitting the ISM of the nearby galaxy triggers a fresh episode of star formation (positive feedback). Such a thing has probably not happened in RAD-18 as the companion is a gas-poor early-type galaxy and no optically blue or UV-bright region is seen at the suspected location of jet-ISM interaction. Jet hitting ISM of another galaxy is an ideal laboratory to investigate details of physical processes in AGN feedback. To the best of our knowledge, this is a unique case, showing two episodes of radio galaxy lobes both during the merger and expanding lobe to have interacted with the merging companion. Several follow up studies can be performed to extract exact dynamical parameters of the interaction. This serves as an ideal laboratory for understanding jet feedback during merger. Independent of assumptions of synchrotron spectral aging we can attempt to find out how long the quiescent phase of nuclear activity can last. The spectral aging values for the duration of quiescent phase, which can be as high as a few tens of Myr (Konar et al. 2013), can be independently verified with future kinematics study and modeling of the merger process.

## 3 Speca: Possible first massive spirals with massive black holes?

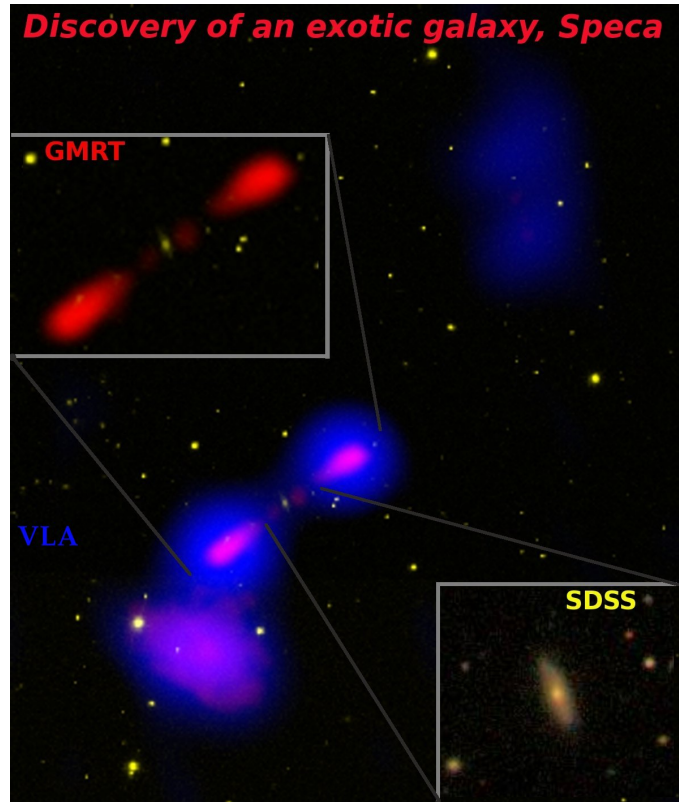

*Fig. 8: Radio-optical image of Speca, based on Hota et al. (2011). Here, red ( GMRT, 325 MHz, b=9"), blue (VLA, NVSS, 1400 MHz b=45") and yellow (SDSS r'-band) show the giant radio source with the disk host (bottom inset from SDSS) and inner double (top inset GMRT+SDSS).*

Merger of two large spiral galaxies and subsequent loss of the cold gaseous medium, through various feedback processes, has been the preferred models for galaxy formation and evolution (Springel, Di Matteo & Hernquist 2005, Hopkins et al. 2006). In the process of merger, galaxies from the `blue-cloud' migrate to the `red-sequence', grow in mass and become increasingly red in colour (Schawinski et al 2007). In the colour-magnitude diagram, lack of galaxies in the region between blue cloud and red-sequence, so called `green valley', suggests a rapid loss of cold gas, fuel for star formation, through dramatic effect of multiple feedback process (Schawinski et al 2014). Furthermore, this is the region where galaxies have been found to be actively accreting. Convincing, "smoking gun" observational evidences are still lacking to demonstrate in any sample that gas-loss by AGN feedback has quenched star formation in post-merger early-type galaxies or in other words in the green valley galaxies. For a cause-and-effect analogy -- the smoke of the gun, assumed responsible for the death of a person by bullet, has to be found older (or of similar time-scale) than the time since death of the person. Current observations find that the age of the relics of AGN activities, both in radio or optical, have been found to be order of magnitude younger than the age of the stars in the galaxy. This intensity weighted age of the stellar population represent the time-scale ever since the star formation has been declining or when in the recent past the galaxy has gone through a major episode of star formation. As the relic radio lobes will be faint, fuzzy and steep spectrum synchrotron plasma, sensitive low frequency observations and visual inspection by citizen-science can detect them most efficiently.

In the process of merger the mass of the galaxy increases and similarly through accretion the mass of the black hole and presumably the spin also increases at least in some cases. Eventually the remnant becomes elliptical and central supermassive black hole launching the large 100 kpc radio jets are always seen associated with elliptical galaxy. Thus, merger has been the standard paradigm to explain why almost always radio galaxies are found hosted in gas-poor elliptical galaxies (Blandford & Zjanek 1977, Heckman et al. 1986, Wilson & Colbert 1995, Sikora, Stawarz & Lasota 2007, Hopkins et al. 2008). Recent discovery of galaxies, however small in number, with large, 100 kpc-several Mpc-size, radio jets with clear star-forming and rotating spiral hosts, suggests processes alternative to merger. The alternate process, however low the probability may be, can create the "central engines" capable of launching massive energy feedback to the host galaxy as well as to the surrounding medium, intra-cluster medium and inter-galactic medium.

Reader is referred to the Kharb et al. (2016) in this journal volume for discussion on these giant spirals hosting giant radio galaxies. Considering all the objects discovered so far we try to subjectively filter out some objects and look for possible general trend in these systems (Ledlow, Owen & Keel 1998, Hota et al. 2011, Bagchi et al. 2014, Mao et al. 2015, Singh et al. 2015, Mulcahy et al. 2016). We find J0315-1906, Speca, J2345-0449, J0836+0532 and MCG+07-47-10 are only five cases of clear spiral-host large radio galaxies, not confused by interaction or small spiral feature buried inside large bulge. Interestingly two out of these five, Speca and J2345-0449, host giant radio lobes with at least two episodes (or, they are DDRG in nature). Speca (Fig. 8) and J2345-0449 are fast rotating massive spirals with rotation speeds reaching ~370 and ~430 km/s, respectively (Speca; Hota et al. 2011, 2014 and J2345-0449; Bagchi et al. 2014 ). Whereas J0315-1906 is clearly not such a fast rotating galaxy and spectroscopic data on J0836+0532 and MCG+07-47-10 are not yet available. This common trend of massive/giant star forming spirals with giant episodic radio lobes, in some sense suggest the opposite of rare red spirals found by citizen scientists. Massive red spirals have been suggested to be the first big spirals formed in the Universe who has been slowly evolving passively, avoiding merger and gas accretion. Here Speca-like objects with star-forming disks and episodic jets, also avoid merger but probably still continue to accrete gas, grow the disk and grow the central black holes. As such galaxies are rare in low-z (where we are able to resolve the optical disk to be clearly spiral) and quasar or radio galaxy formation peaked at redshift of 2-3 (Nesvadba et al. 2008 ), we may expect future study to find many such galaxies with higher than the redshift of Speca (0.137) which is the highest among the all such objects. We speculate that at higher redshift where massive spirals and luminous quasars were more prevalent than ellipticals, Speca-like galaxies could be more common and high resolution optical and sensitive low-frequency SKA observations may be helpful. The first supermassive black holes formed in the Universe are currently residing at the centres of massive clusters (McConnell et al. 2011, Wang et al. 2015). These brightest cluster galaxies, hosting supermassive black holes of billion solar mass, have been modified so much by frequent interactions that they no longer carry any signature of their earliest nature, like spiral arm, bar, psudo-bulge, KDC, dual-nuclei etc.. However Red spirals and Speca-like radio galaxies may still contain signs of their earliest behaviour like "vestigial organs". Indeed, existance of psudo-bulge in J2345-0449 and newly discovered, gamma-ray emitting, radio loud, narrow-line Seyfert 1 galaxy PKS 2004-447 firmly support secular evolutionary processes (not merger-driven) for these rare systems (Oshlack, Webster & Whiting (2001), Kormendy & Kennicutt (2004), Abdo et al. 2009, Kotilainen et al, 2016). Thus, Speca-like galaxies may indeed be the "living fossil" or opportunity to learn about first massive spirals formed with massive black holes in their centres in the early Universe.

**4. Citizen-science Galaxy Zoo and Radio galaxy Zoo:** The visual classification of galaxies into categories like spiral, bar-spiral, elliptical, merger, irregular, clockwise vs. anti-clockwise spirals etc. is very simple and can be done even by a high school student. But this becomes tedious to do for a large number of galaxies for any professional astronomer. Automated computer algorithms can be developed for such pattern recognition tasks to handle digitised images of a large number of galaxies. However, for faint, fuzzy and morphologically complex sources, a combination of human eye and brain performs better than machine learning processes (e.g. Jain et al. 2000). In the year 2007, Kavin Sahawinski and Chrish Lintott created an interactive website Galaxy Zoo (www.galaxyzoo.org) where people on the internet can participate in such simple galaxy morphology classification. The interface displays different image to different person on the web and every click takes you to a different galaxy. Classification of a single galaxy can be confirmed by numerous members before professionals look into the results. Immediately after launching of the website, thousands of people joined and started classification of galaxies. During on-line discussion of these galaxies they discuss strange unexpected objects. They also efficiently identify special type of objects astronomers have asked them to identify. This way it solves the problem of the astronomer, tedious job of going through a large number of galaxy images, by distributing the herculean task to a large number of web-based volunteers. They not only help classify but also help discover unexpected galaxy types and help grow a sample of special types of galaxies, as asked by the astronomers, by searching in the large data bases.

With the combination of people's power, computing power, and internet available at everybody's finger tip, very soon amazing discoveries were made. As mentioned above, a faint fuzzy object blue in the SDSS colour image was spotted by a Dutch school teacher. As it did not look anything like galaxies, she has seen before, she asked in the on-line group discussion, if anyone knew what that blue blob of faint fuzzy emission is. Professional astronomers were also stunned to find such a strange object, not seen in any other bands of SDSS except one, suggesting it to be emission line gas cloud. With follow up observations this finding was published and the school-teacher citizen-scientist, the discoverer, Hanny van Arkel, was the fourth author of the discovery paper (Lintott et al. 2009). Citizen scientists were asked to spot similar objects and they reported dozen such new objects to professional astronomers. This has been followed up with the HST and many other telescopes and was published by Keel et al (2012). In that paper, one of the citizen scientists Richard Proctor, a telecommunication consultant, has been awarded a co-authorship. This finding of AGN-ionised clouds has indeed opened up a new window to investigate the history of AGN/quasar activity.

Another such stunning discovery from Galaxy Zoo is Red Spirals. Usually spirals are gas-rich young star-forming systems and appear blue in optical colour images. But, these rare spiral galaxies are relatively red in colour suggesting that they have been passively evolving since very long time consuming its fuel for star formation slowly (Masters et al. 2010). They could also be the very first big spirals formed in the Universe. No radio continuum study on such objects, to look for past AGN activity, has been performed yet. We have been looking at these sources in the TGSS data. Similarly "Green peas" galaxies is another surprise from citizen science in astronomy (Cardamone et al. 2009). They are extremely compact and star bursting galaxies. The bright and large equivalent width of [O III] Oxygen emission line (5007 Angstrom) from these unresolved galaxies make them look green in the SDSS colour images. These have the highest specific star formation rate ($10^{-8}$ per year) in the local Universe. They serve as local laboratory for high-z galaxies like Lyman Break Galaxies and Lyman α emitters.

GMRT has been used to detect Green Peas galaxies. Combined with data from VLA they were found to have extremely high magnetic field comparable to or larger than 30 µG, which is larger than that of our Milky Way Galaxy (Chakraborti et al. 2012). They stated, "This is at odds with the present understanding of magnetic field growth based on amplification of seed fields by dynamo action over a galaxy's lifetime. Our observations strongly favour models with pre-galactic magnetic fields at µG levels". Green Peas galaxies are considered ideal laboratories to understand the basics of star formation in the early Universe.

On 17th December 2013, Galaxy Zoo expanded its project to radio wave length and named it Radio Galaxy Zoo (https://radio.galaxyzoo.org/). The website asks public to identify the host galaxy of the radio source by clicking on the right position in an image being displayed with radio-infrared overlay. Radio images from FIRST survey using VLA and Australia Telescope Large Area Survey (ATLAS) using Australia Telescope Compact Array (ATCA) were superposed with infrared images from Wide-Field Infrared Survey Explorer (WISE) or Spitzer Space Telescope. Participants can discuss interesting or confusing cases further in the online discussion forum. The host-galaxy identification and radio source classifications have helped investigate radio-infrared nature of galaxies in a large galaxy sample (Banfield et al. 2015). Recently a poor cluster of galaxies (RGZ-CL J0823.2+0333) has also been discovered by the Radio Galaxy Zoo. This cluster was traced by a giant (1.1 Mpc) wide angle tail radio galaxy named RGZ J082312.9+033301. What is unique in this source is, it is wide angle tail and episodic in nature, a clear double-double radio galaxy (possibly triple-double), breaking all rules of double-double lobe formation mechanism. The finding has been reported by Banfield et al. (2016) with both the citizen-scientist discoverers from Russia, Ivan A. Terentev and Tim Matorny, as co-authors in it. Selecting from the list of host galaxies of the radio sources, identified by these citizen-scientists, an extremely rare, spiral-host radio galaxy, similar to Speca, was also discovered by Mao et al. (2015). It is to be noted that Galaxy Zoo is now a part of the bigger association named Zooniverse which runs citizen science projects not just in extragalactic astronomy but in almost all branches of science e.g. space, climate, medicine, humanity, nature. Till the writing of this paper Zooniverse (www.zooniverse.org) has published 108 research papers and out of which 48 publications have come from extragalactic research by Galaxy Zoo. Recently, an excellent review on citizen-science has been published by Marshall, Lintott & Fletcher (2015). This review discusses its history and promise of its power for future astronomy education, research, and social development. For even broader social implications of citizen-science reader is refered to the book titled "Reinventing Discovery: The New Era of Networked Science" by Nielsen (2011)

**5. RAD@home modified citizen-science:**
Soon after the discovery of Speca (Hota et al. 2011), we realised that it could have been discovered by anybody just using decades-old DSS in optical, and NVSS (Condon et al. (1998)) and FIRST ( Baker et al. 1995) in the radio. All it requires is little training and publicly available web-tool to overlay radio and optical images. Staying connected through Internet and a web-portal to share images and discuss them in collaboration with a professional is enough for large number of citizens to discover such rare radio galaxies. In concurrent with this discovery, TGSS sky survey from GMRT at 150 MHz was being released and we realised the golden opportunity. A process of training citizens who have basic undergraduate science education is needed for full exploitation of the TGSS data and discover not only Speca-like rare objects but all kinds of objects that are important to professional astronomers. The follow up observations of those discovered objects with the GMRT and VLA are also required. The approach taken by Galaxy Zoo project indeed helps professionals extract interesting patterns from huge databases. With multiple investigation of the same target by different citizen-scientists, professionals reduce their time on analysing the mistakes or false-reports by these volunteers. This solves the problem of big-data for the professionals, but, unless it is a single-object big discovery citizen-scientists don't really get authorship credit and cannot grow much in their personal career. Along with solving the problem of the scientists we wished to solve the problem of the people who are passionate about astronomy. People in remote locations, away from astronomy research institutes, hardly get the exposure to excitements in astronomical research. Public lectures, public outreach programs and participation in undergraduate projects are the typical ways to inform, inspire and involve students. The geographical non-uniformity of astronomical research institutes in India naturally leads to an unequal opportunity in accessing those national facilities when we consider people not pursuing PhD. One powerful way to address such problems is the use of Internet. Internet has the transformational power which can involve a million people, like in the Galaxy Zoo project, to solve large-scale problems of the society. A modification to existing citizen-science research was needed so that through web-based selection and web-based training, participants even in remote locations get an equal opportunity to grow academically. Making education and research accessible through Internet can not only provide equal opportunity for growth to all citizens but also naturally alleviate many socio-economic and geo-political constraints on inclusive growth, especially faced by remote under-developed regions. Hence a modified citizen-science programme or an inter-institutional e-education e-research project, was proposed to various institutes in India.

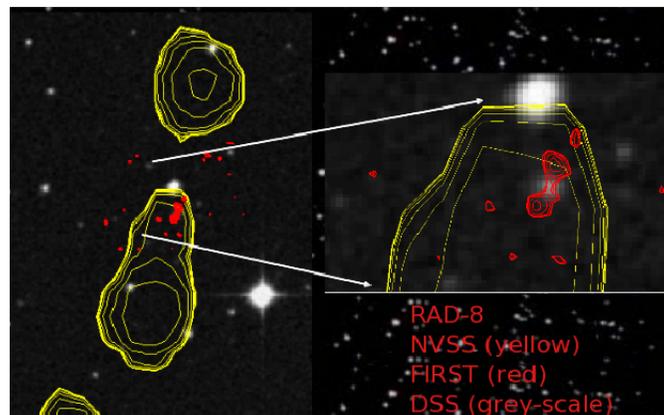

*Fig. 9: A new double-double radio galaxy RAD-8 found by RAD@home citizen-scientists as reported briefly by Hota et al. (2014).*

In April 2013, RAD@home the first Indian Citizen-science research project in astronomy, a zero-funded and zero-infrastructure initiative, was launched in Google and Facebook. Real Astronomy Discovery sitting at home anywhere in India was abbreviated to RAD@home. The appeal in Facebook asked science-educated citizens, anybody who has entered University education (any BSc/BE) irrespective of their status, student/employed/unemployed, to join the Facebook group (https://www.facebook.com/groups/RADathome/). The tag line, "Any BSc/BE Can Do research" using GMRT data, was hash-tagged as #ABCDresearch and this helped the message reach wide population. Currently, the group has over 1800 members. Newly joined group members were given step-by-step instructions to make UV-Optical-IR-radio multi-wavelength Red-Green-Blue RGB-colour images of galaxies and radio galaxies using NASA Skyview (http://skyview.gsfc.nasa.gov). These images were shared and discussed by all participating members in the group. Members who were found good, were declared RGB-qualified and further called to get trained in a week-long RAD@home Discovery Camp. Camps have been hosted by various research institutions like Institute of Physics (Bhubaneswar), Harishchandra Research Institute (Allahabad), UM-DAE Centre for Excellence in Basic Sciences (Mumbai). The camp in Delhi was hosted by astronomy outreach organisation Nehru Planetarium (New Delhi) and supported by Vigyan Prasar (Govt. of India, New Delhi). During the camp, in addition to normal astronomy lectures, citizen-scientists discover potentially new radio sources from the TIFR GMRT Sky Survey (TGSS) images. As TGSS images at 150 MHz have the best resolution and sensitivity at such a low radio frequencies, in every camp they have discovered cosmic sources that is worth follow up investigation by professional astronomers. GMRT Time allocation Committee (GTAC) has awarded observing time, going through standard international competition, for three cycles in a row. This program named, GOOD-RAC: GMRT Observation of Objects Discovered by RAD@home Astronomy Collaboratory, has been appreciated by SKA-India Consortium as a successful model. RAD@home was discussed as a unique model in the international review paper on citizen-science in astronomy by Marshall, Lintott & Fletcher (2015). So far, the total number of e-astronomers trained in these camps is 69. They are distributed all over India, from all walks of life and continue to work from home at some scheduled time on-line and/or at their convenient time off-line. They work mostly in two modes. 1. Discover new and interesting sources in the large 4.5 degree FITS files of TGSS survey images using SAO ds9 image analysis software, 2. Analyse assigned target galaxies using the widely available multi-wavelength data which when combined from many members make it a detail analysis of a specific sample of galaxies. They share the images they generate (without disclosing RA and Dec of their targets) and discuss it with fellow e-astronomers. They generate their pre-discovery report documents in Google-doc known as the First Investigation Report (FIR). This is facilitated with regular online e-class, three hours a week (Wednesday morning and evening, and Sunday morning). These e-classes are basically Facebook group-chats discussing educational video, multi-wavelength image analysis of the new found objects from TGSS, analysis of newly discovered objects and associated publications if any. One can participate in this fixed-time fixed-day e-class from anywhere with even cellphone with internet. People having problem with the time-schedule can post their new-found galaxy images, without disclosing the RA Dec (to protect credit of discovery but still learn by discussion), in the e-astronomer Facebook group containing all 69 camp-trained citizen-scientists. Observational results so obtained by these e-astronomers are presented in national and international conferences with discoverers as co-authors in the paper.

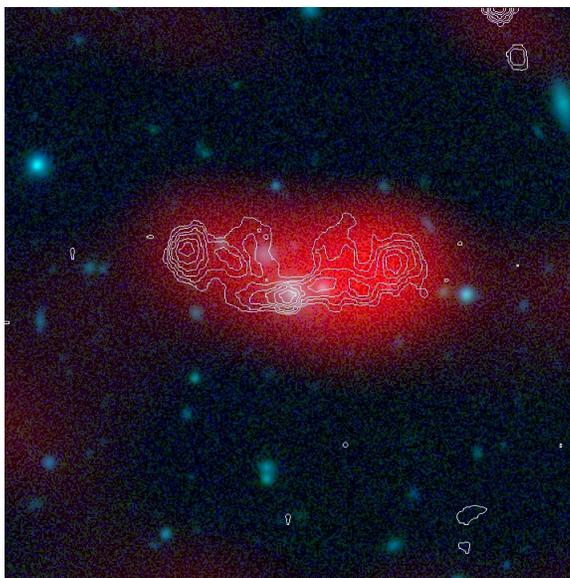

*Fig. 10: As briefly reported in Hota et al. (2014), RGB-C image of bent-lobe radio galaxy RAD-16. Here, SDSS optical image is shown in green+blue, red shows 150 MHz image (GMRT, TGSS ADR1), and contour shows 1400 MHz (VLA, FIRST). Note that the 2 Mpc long galaxy filament extends north-south, consistent with the bending.*

Some of our interesting results were briefly presented for the first time in the international conference Meterwalenegth Sky organised to celebrate 10 years for GMRT as an international research facility (Hota et al. 2014). Our results include, several new candidate episodic radio galaxies, candidate Speca-like spiral-host star forming radio galaxies, bent-lobe radio galaxies residing in new galaxy filaments, relic or dead-lobe radio galaxies, and relic/halo diffuse radio emission in clusters of galaxies. The list of each type of objects has been growing steadily and with the release of TGSS ADR (Intema et al 2016) the scope has increased ten fold. A journal paper compiling all these is in preparation (Hota et al. 2016). Some of those findings are presented here.

RAD-8 (RA 11 57 36.0, Dec -03 14 41) is a beautiful example of episodic galaxy that the citizen-scientists found, where the inner lobe is clearly seen in the VLA FIRST and outer lobes are visible only to NVSS. Interestingly, this source is not detected in TGSS. Many candidate DDRGs have been observed with the GMRT under the GOOD-RAC project. RAD-16 (RA 11 45 26.0, Dec -02 23 33 ) is a bent-lobe radio galaxy which is sitting in the middle of a nearly 2 Mpc long filament of galaxies. Only one such clear case (Edwards, Fadda & Frayer 2010) is known in the literature and further study in this line is in progress. Such objects are helpful in determining the density of gas in the intra-filament medium which is warm and so thin that its emission is undetectable to any wavelength of observation. RAD-18 is a rare case of episodic radio galaxy hosted in an early-type-early-type dry merger where one lobe in one episode has interacted with the merging companion. Details of this has already been presented in Sec 2.3.3.

## 6. Future Plans for science with SKA, uGMRT and RAD@home.

Our plan is to take lead from our continuing low-frequency studies with the GMRT. It has basically two approaches. 1. Discovery of missing-link rare objects and expanding the sample size of such objects via multi-wavelength search by large community of citizen-scientists on line. 2. The second approach is detailed multi-wavelength observation of important targets in collaboration with professionals from all over the world and using all possible telescopes. The approach of RAD@home, a modified citizen-science project, can fit into almost any targeted science proposal submitted to SKA. To understand this, a few critical differences between typical citizen-science projects like Radio Galaxy Zoo and RAD@home is worth discussing. The fundamental difference is the foundation on which our plan with SKA is building upon. As stated earlier, standard citizen-science solves the big-data problem faced by the astronomers. Although currently at a much smaller scale, RAD@home also attempts the big-data problem. In addition, it also addresses the problems being faced by astronomy-passionate students and citizens who already have undergraduate level science education. RAD@home follows a direct face-to-face interactive research training for a minimum of one week, as well as a regular on-line e-class. It is a directly trained on-line work-force for radio-astronomy research. Not only many of the students become co-investigator (Co-I) in the GMRT proposals, they also get a boost in their career in getting internship or PhD studentships in astronomy research institutes. The students undergo a selection process to get the interactive research training and a recommendation letter to join research institutes based on their performances. The sky fields are distributed without much overlap, which ensures some success to each and every sincere e-astronomer. It is much like sharing resources for inclusive growth and avoiding conflict of interest. This way RAD@home can alleviate infrastructure constraints of research institutes like hostel rooms, number of students with fellowship etc. This can expand the user community or human-resource network of national research facilities (telescope) to socio-economically and geo-politically backward regions through the internet. Imaging data from uGMRT, or in future SKA, will be a world-class research resource which can provide equal opportunity of growth through participation in research to all people irrespective of their location, city or town, economic status- poor or rich etc.. Currently, it is TGSS data but soon larger amount of uGMRT imaging data, through GOOD-RAC, can be created and provided to these citizen-scientist e-astronomers spread all over the nation looking for opportunity of career growth and/or simply to have intellectual happiness in exploring galaxies and black holes in the cosmos.

As uGMRT will be like one tenth of SKA in collecting area, many pilot studies can be performed with low-z examples using uGMRT before expanding to high-z cases with the incredible sensitivity of SKA. Not only the sensitivity but also the field of view of SKA will be significantly high, compared to GMRT-like existing interferometers. If 15 m dishes are used, it will be nearly ten times bigger than that of GMRT. This aspect has often been ignored in many proposals with targeted-science cases submitted to GMRT and being planned for SKA by radio continuum science working groups. Take for example the study of giant radio galaxies, which incorporate both GMRT and VLA, which are complimentary in frequency range and are really needed for spectral ageing study of giant as well as double double radio galaxies. In the sample of ten giant radio galaxies studied by Konar et al. (2008), the angular size of the target-of-interest sources are very small (6.16, 6.25, 5.75, 3.82, 6.46, 11.4, 3.33, 2.96, 3.26, 3.16 arc min) compared to the GMRT field of view of 186, 114, 81, 43, and 24 arc min for 151, 235, 325, 610, and 1420 MHz bands respectively. These numbers suggest that we often barely use a few percent (6 x 6 of 80 x 80 that is 0.5 % for 325 MHz of GMRT) of imaging data. The remaining equally sensitive images remain in the hard disk of the scientists forever or are published just as catalogues. When we consider the study of general galaxy population, this is an extremely inefficient use of telescope time unless the full field of view of an observation is utilised in an organised way to extract all possible science incorporating multi-wavelength data freely available on the web. This is where citizen-science can play an efficient role, looking at each and every galaxy in the field and help the professionals in increasing their publication record. Through its Facebook page (https://www.facebook.com/RADatHomeIndia/) RAD@home is planning to launch a program named Donate FITS Images for Recycle (DFIR) where the PI's of GMRT proposals will be requested to share the FITS image files, which has already been used in their publications. PI and raw data analyst will get due credit in the RAD@home publications. There is definite benefit of these files in the education and discovery through citizen-science approach.

We wish to follow this approach in uGMRT and SKA era. We will propose target galaxies of high scientific justification and the full field of view of the SKA image will be supplied to citizen-scientists for multi-wavelength investigation using all the available data on the internet. Characterising diffuse faint emission in multi-frequency and angular-scale-sensitive interferometric images can be done much better by a trained-human (e-astronomers) than any computer algorithm. In this aspect interferometric images are very complex compared to optical colour images. For example a diffuse emission blob seen bright in low resolution NVSS can be missing in higher resolution VLA FIRST, although both are at the same wavelength and FIRST is more sensitive (nearly three times less r.m.s. noise) than NVSS. This is what is the first clue to finding diffuse emission as relic radio lobes and radio relics in clusters of galaxies. Over a dozen such examples, missing in high resolution images but visible to low resolution images, have been identified by RAD@home citizen scientists. SKA observations are likely to provide direct FITS images and can easily be analysed by citizen-scientist e-astronomers of RAD@home sitting at home. This is why it is aptly named Real/Rapid Astronomy Discovery sitting at home anywhere in India, by-passing the need of travel to research institutes to participate in research. This is how a resource-poor citizen can be connected with cutting-edge research projects, being stationed in a small backward town just by using Internet, and enrich his/her intellectual property and contribute in the process of the nation-building. We do have proposals to reward them in a way that will be meaningful to their local life and not just with international publications. This modified citizen-science with uGMRT and SKA can indeed use astronomy as a medium for development. Future sky seems bright irrespective of the wavelength we look at.

*Acknowledgment:* AH is thankful to University Grants Commission (India) for the one-time start-up and monthly salary grants, under the Faculty Recharge Programme. We are thankful to Institute of Physics (Bhubaneswar, India), Abdus Salam International Centre for Theoratical Physics (ICTP, Italy), India chapter of the International Union of Radio Sciences (URSI), Tata Institute of Fundamental Research (TIFR), Homi Bhabha Centre for Science Education (HBCSE-TIFR), International Centre for Theoratical Sciences (ICTS-TIFR), National Centre for Radio Astrophysics (NCRA-TIFR) and Astronomical Society of India (ASI) for appreciation and supports to present our results in various conferences and help the citizen-science grow in India. National Radio Astronomy Observatory (NRAO) of the National Science Foundation (NSF) and, California Institute of Technology (Cal-Tech), Jet Propulsion Laboratory (JPL) and National Auronotics and Space Administration (NASA) have been acknowledged for the press releases related to Speca and NGC3801, respectively. We thank Prof. Govind Swarup, Prof. S. Ananthakrishnan, Prof. Ajit K. Srivastava, Prof Sudhakar Panda, Dr. Sam Pitroda, Prof. Ashoke Sen, Prof. S M. Chitre, Dr. K. Kasturirangan, Prof Ramaswamy Subramanian, Prof Swadhin Pattnaik, Dr. N. Rathnasree, Dr. Arvind Ranade, Prof. Willem Baan, Prof Jocelyn Bell Burnell, Dr. Judith H. Croston, Prof. Katherine Blundell for their support to the project RAD@home ( #RADatHomeIndia #ABCDresearch ).